\documentclass[final]{siamltex}

\usepackage{graphicx}
\usepackage{amssymb}
\usepackage{subfigure}

\def\v{\mathbf}

\title{A New Multilevel Method for Electrostatic Problems Through Hierarchical Loop Basis}

\author{Zu-Hui Ma \thanks{Department of Electrical and Electronic Engineering, The University of Hong Kong, Hong Kong; Department of Electrical and Computer Engineering, University of Illinois at Urbana-Champaign, Urbana, IL 61801 USA. ({\tt mazuhui@hku.hk}). }
        \and Weng Cho Chew \thanks{Corresponding author. Department of Electrical and Computer Engineering, University of Illinois at Urbana-Champaign, Urbana, IL 61801 USA ({\tt w-chew@uiuc.edu}).}
        \and Yu Mao Wu \thanks{Department of Electrical and Electronic Engineering, The University of Hong Kong, Hong Kong; Department of Electrical and Computer Engineering, University of Illinois at Urbana-Champaign, Urbana, IL 61801 USA. ({\tt ymwu@eee.hku.hk}). }
        \and Li Jun Jiang \thanks{Department of Electrical and Electronic Engineering, The University of Hong Kong, Hong Kong.({\tt jianglj@hku.hk}). }}

\begin{document}

\maketitle

%\renewcommand{\thefootnote}{\fnsymbol{footnote}}
%
%\footnotetext[1]{Department of Electrical and Electronic Engineering, The University of Hong Kong, Hong Kong.}
%\footnotetext[2]{Department of Electrical and Computer Engineering, University of Illinois at Urbana-Champaign, Urbana, IL 61801 USA.}
%\footnotetext[3]{Corresponding author. Department of Electrical and Computer Engineering, University of Illinois at Urbana-Champaign, Urbana, IL 61801 USA (w-chew@uiuc.edu).}

%\renewcommand{\thefootnote}{\arabic{footnote}}

\begin{abstract}
We present a new multilevel method for calculating Poisson's equation, which often arises form electrostatic problems, by using hierarchical loop bases. This method, termed hierarchical Loop basis Poisson Solver (hieLPS), extends previous Poisson solver through loop-tree basis to a multilevel mesh. In this method, Poisson's equation is solved by a two-step procedure: First, the electric flux is found by using loop-tree basis based on Helmholtz decomposition of field;  Second, the potential distribution is solved rapidly with a fast solution of $O(N)$ complexity. Among the solution procedures, finding the loop part of electric flux is the most critical part and dominates the computational effort. To expedite this part's  convergent speed, we propose to use hierarchical loop bases to construct a multilevel system. As a result, the whole solution time has been noticeably reduced. Numerical examples are presented to demonstrate the efficiency of the proposed method.
\end{abstract}


%\begin{AMS}
%15A15, 15A09, 15A23
%\end{AMS}

\pagestyle{myheadings}
\thispagestyle{plain}
\markboth{Z.-H. MA, W. C. CHEW, Y. M. WU, AND L. J. JIANG}{NEW MULTILEVEL POISSON SOLVER}

\section{Introduction}
Numerical solutions of Poisson's equation have been found to be of great importance in various scientific and engineering problems, such as nanodevice design, fluid dynamics, and electrochemistry~\cite{FeynmanLecture2,Datta_book,ChengLiu2007}.

\begin{figure}[h]
\centerline{\includegraphics[scale =0.4]{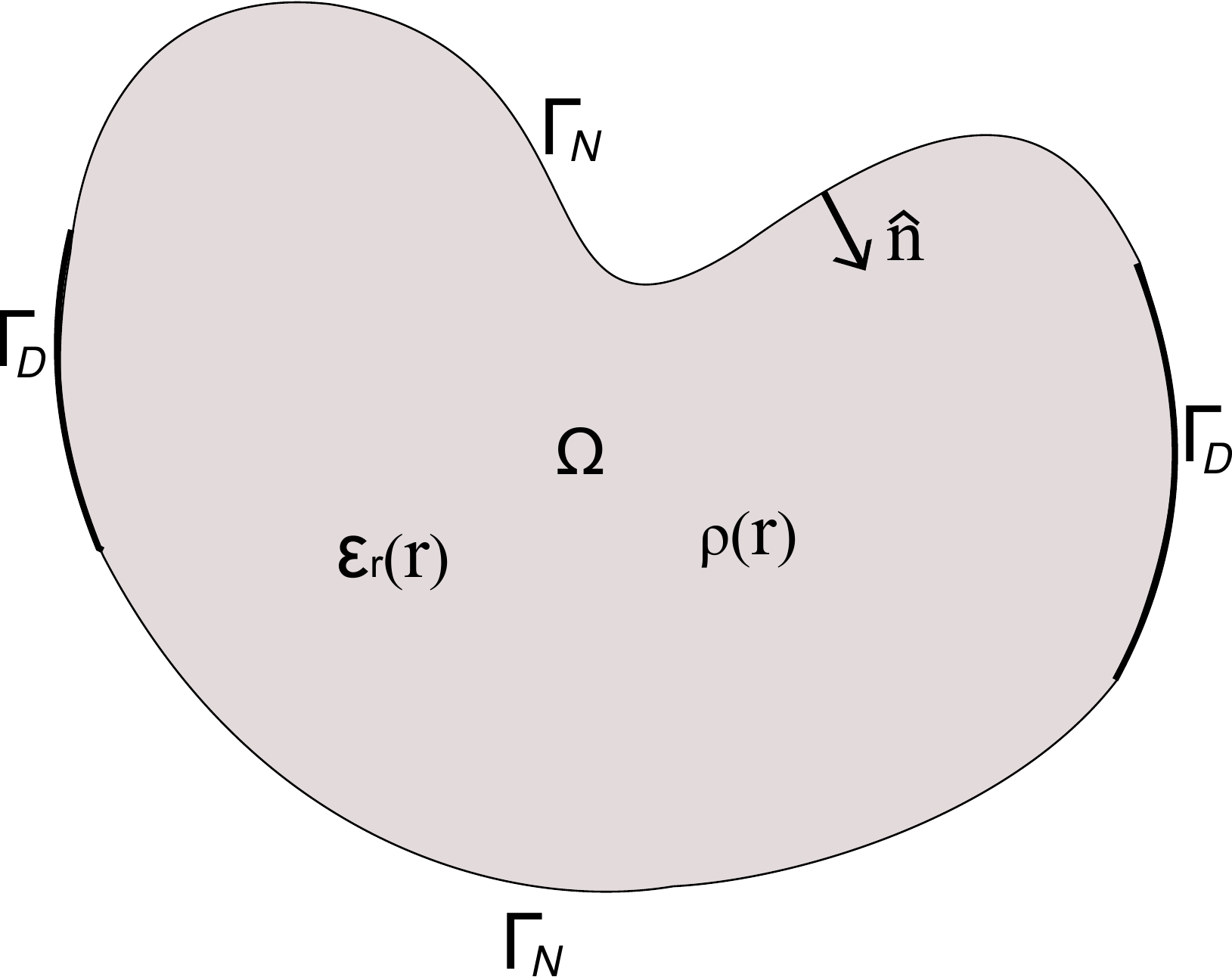}}
\caption{Schema for a typical Poisson problem.} \label{fig:PE}
\end{figure}

In electrostatics, when a simply connected region $\Omega$ is occupied by inhomogeneous dielectric materials as shown in Fig.~\ref{fig:PE},  the corresponding problem is governed by the following equations
\begin{equation}\label{eq1.1}
\begin{array}{l}
\nabla\times\mathbf{E}(\mathbf{r})= 0 \\
\nabla\cdot\mathbf{D}(\mathbf{r})= \rho(\mathbf{r}),
\end{array}
\end{equation}
where $\mathbf{E}(\mathbf{r})$, $\mathbf{D}(\mathbf{r})$ denote the electric field and the electric flux, respectively, and $\rho(\mathbf{r})$ is the electric charge density. Under the assumption of linear, isotropic media, the electric flux $\mathbf{D}(\mathbf{r})$ relates the electric field by
\begin{equation}\label{eq1.2}
\mathbf D (\mathbf r)= \epsilon(\mathbf r)\mathbf{E}(\mathbf{r}),
\end{equation}
where the permittivity $\epsilon(\mathbf r)= \epsilon_0 \epsilon_r(\mathbf r)$.  $\epsilon_0$ is the permittivity of free space, while the relative permittivity $\epsilon_r(\mathbf r)$ is position dependent generally.
By introducing the electrostatic scalar potential in Eq.~(\ref{eq1.1}) such that $\mathbf E=-\nabla\phi$,
we have Poisson's equation as follows
\begin{equation}\label{eq1.3}
\nabla\cdot \left(\epsilon_r(\mathbf{r})\nabla\phi(\mathbf{r})\right)=-\frac{\rho(\mathbf{r})}{\epsilon_0}.
\end{equation}
The paramount task is to obtain solutions of the above with boundary conditions of corresponding problems.

At the present time, existing numerical methods for Poisson's equation are grouped into two categories: direct and iterative solvers. Within direct methods, the multifrontal method is one of the most efficient algorithms. In~\cite{Xia2010},  a superfast multifrontal method has been developed to take advantage of hierarchical tree structures of both hierarchically semiseparable (HSS) matrices and the classical multifrontal idea. It leads to a total complexity of $O(N^2)$. However, $O(N^2)$ is still unacceptable for large problems. As for iterative solvers, one competitive method is based on the fast multipole method
(FMM) integral equation scheme~\cite{McKenney1995,HuangGreengard2000,Ethridge2001}. Alternatively, the multigrid method~\cite{MG_book,MG_rev,MGtutorial2000} is the most popular one because it could achieve nearly optimal complexity in theory.

Recently, a novel Poisson solver has been proposed to solve 2D problems \cite{fpsnbc,myJCP}. In contrast to traditional Poisson solvers, this method solves the electric flux $\mathbf D$ directly. The
electric flux is expressed as the combination of the loop space
(subspace) (solenoidal or divergence free) part and the tree space
(subspace) (quasi-irrotational) part. These two spaces, however,
are non-orthogonal to each other that is dissimilar to the rigorous
Helmholtz decomposition. By expanding the electric flux density vector $\mathbf{D} (\mathbf r)$ (denoted by electric flux for short throughout) with two sets of basis functions: loop
and tree basis functions, the electric flux can be solved by a
two-stage process: First, to find the tree-space part, a matrix
system is derived based on
$\nabla\cdot\mathbf{D}=\rho$, and then it is solved by a fast direct tree
solver with $\textit{O}(N_t)$ complexity, where $N_t$ is the total
number of tree basis functions.  The obtained electric flux is
nonunique since a divergence free component is part of its null space. Second, the loop-space part of $\mathbf E$ is acquired by a projection
procedure, which is iterative.
Once the electric field is obtained, we can readily get the
potential distribution by solving $\mathbf E=-\nabla \phi$ by the
fast tree solver as well.  This method affords a new way to solve Poisson's equation that is faster than the traditional finite element methods (FEM). Moreover, almost linear complexity has been observed when stopping criterion is not less than $1\times10^{-3}$. However, the solution time could deteriorate as more accurate results are required.

To enhance the efficiency of this method,  a method based on multilevel analysis of differential operators provides a good option. One important multilevel approach is the hierarchical linear Lagrangian basis (nodal basis) method that was proposed by Yserentant about two decades ago \cite{Yserentant1986,Yserentant1986347}. In this method, the FEM basis is changed from a single-level one to a multilevel basis that spans the same space. Deuflhard \textit{et al.} soon afterwards reported an adaptive multilevel FEM code \cite{Deuflhard1989}. For this method, it can achieve the same kind of computational complexity without use of standard multigrid techniques. This kind of methods are also regarded as Hierarchical basis preconditioners (HBs) and applications in three dimensions are given  in~\cite{Ong_sisc1997,Bornemann1993}. In \cite{Vipiana2005,AndriulliSisc}, a hierarchical vector-valued basis on triangular mesh has been proposed to solve electric field integral equation (EFIE) with method of moment (MoM). This basis can be further decomposed into solenoidal part and irrotational part; The solenoidal part comprises hierarchical loop basis that has close relation with hierarchical nodal basis.

Another important category of multilevel methods are those based on wavelet theory. In the last few decades, wavelet methods \cite{Daubechies1988,Strang1989} has been developed as a powerful tool in numerous areas of mathematics, engineering, computer science, statistics, physics, etc. In the early stage, traditional wavelets consist of scaled and shifted versions of a single function on a regularly spaced grid. Sweldens~\cite{Sweldens1996} break this restriction by proposing lifting scheme, which led to more wider class of second-generation wavelet. Then, the hierarchical loop basis, in view of~\cite{Vipiana2005}, could be considered as a special kind of wavelet functions.

In this paper, we propose to extend our previous loop-tree based Poisson solver to a multilevel method by using the hierarchical loop basis that has been used for EFIE before. It can speed up the iteration process and then reduce the solution time of our Poisson solver. As compared with multilevel multigrid method, this method is simpler because it is independent of any uniformity restriction on the applied meshes. In addition, this new method is more friendly to parallel computing since all computations are local.

The organization of this paper is as follows. In
Section 2, we derive the formulation and introduce the relative basis functions and vector space decomposition theorem. In Section 3, we briefly outline the algorithm of previous Poisson solver that use normal loop-tree bases. Next, a hierarchical loop basis function is presented in Section 4.  Finally, in Section 5,
we will validate the method and illustrate the efficiency of the
new method. Conclusions will be drawn
in Section 6.

%%%%%%%%%%%%%%%%%%%%%%%%%%%%%%%%%%%%%%%%%%%%%%%%%%%%%%%%%%%%%%%%
%%%%%%%%%%%%%%%%%%%%%%%%%%%%%%%%%%%%%%%%%%%%%%%%%%%%%%%%%%%%%%%%%%%%
%%%%%%%%%%%%%%%%%%%%%%%%%%%%%%%%%%%%%%%%%%%%%%%%%%%%%%%%%%%%%%%%

\section{Preliminaries}
In this section, we define the Poisson problem of interest and introduce some preliminaries that will be used in latter sections.

To define the well-posed problem, we first study the Sobolev space
of the electric field ${\bf E}$ and electric flux  ${\bf D}$ in Eq.
(\ref{eq1.1}). We consider the Lipschitz domain $\Omega$ with the
Lipschitz boundary, and  introduce the following two Sobolev spaces
\begin{eqnarray}\label{HDIV}{\bf H}(\mbox{div,}
\Omega)&=&\{{\bf f}\,|\,{\bf f}\in
\left(L^2(\Omega)\right)^3, \nabla \cdot {\bf f} \in L^2(\Omega)\},\\
\label{HCURLDIV}{\bf H}(\mbox{curl, div,} \Omega)&=&\{{\bf
f}\,|\,{\bf f}\in\left(L^2(\Omega)\right)^3, \nabla \times {\bf f}
\in \left(L^2(\Omega)\right)^3, \nabla \cdot {\bf f} \in
L^2(\Omega)\}.\end{eqnarray} Here, the function ${\bf f}$ is a
vector function in 3D space.

By Eq. (\ref{eq1.1}),  the electric field ${\bf E}$ is curl-free.
Once the charge $\rho(\mathbf{r})$ li es in $L^2(\Omega)$,  ${\bf E}$
and  ${\bf D}$  are curl and  divergence bounded function in the
sense of $\|.\|_2$ norm. Hence, we have ${\bf
E}\in {\bf H}(\mbox{curl,} \Omega)$ and ${\bf D}\in {\bf
H}(\mbox{div,} \Omega)$. Furthermore, for the homogeneous medium,
$\nabla \cdot {\bf D}=\rho$ implies ${\bf E}\in {\bf H}(\mbox{div,}
\Omega)$. In this situation, we have {${\bf E}\in {\bf
H}(\mbox{curl, div,} \Omega)$}

\subsection{Poisson Problem}

The Poisson problem with mixed boundary condition of interest in this paper is as follows:
\begin{equation}\label{s2eq1}
\renewcommand{\arraystretch}{1.5}
\begin{array}{ll}
\nabla\cdot\left(\epsilon_r(\mathbf{r})\nabla\phi(\mathbf{r})\right)=-\rho(\mathbf{r})/\epsilon_0, &\mbox{ $\mathbf{r}\in\Omega$},   \\
\phi(\mathbf{r})=\phi_0(\mathbf{r}), &\mbox{  $\mathbf{r}\in\Gamma_D$},  \\
\frac{\partial}{\partial n}\phi(\mathbf{r})=g(\mathbf{r}), &\mbox{ $\mathbf{r}\in\Gamma_N$}.
\end{array}
\end{equation}
The simulation domain $\Omega\subset \mathbb{R}^2$ is a two
dimensional bounded region, with
boundary $\Gamma$ and normal $\hat{\mathbf{n}}$ that points to the
solution region as shown in Fig.~\ref{fig:PE}. The boundary $\Gamma=\Gamma_D\bigcup\Gamma_N$
is composed of two parts: the first one, denoted by $\Gamma_D$, is
imposed by Dirichlet boundary condition and the other part,
$\Gamma_N$, is imposed by Neumann boundary condition.
Suppose that the
Dirichlet boundary consists of finite $M$ distinct boundaries,
$\Gamma_D= \bigcup_{i=1}^{M}\Gamma_D^{(i)}$, then a fixed potential
$\phi_0(\v r)$ is prescribed on the boundary $\Gamma_D^{(i)}$ for
$i=1,2,...,M$.

To complete the description of a well-posed
problem, the potential $\phi(\mathbf{r})$ belongs to  the
Sobolev space $H^1(\Omega)$, with the super index ``1" as the first
derivative in the weak form sense. The charge $\rho(\mathbf{r})$ and the Neumann boundary data $g(\mathbf{r})$ must be a square
integrable function over the corresponding boundary
\cite{Brown1994}.
In other words, we have $\rho(\mathbf{r})\in
L^2(\Omega)$ and $g(\mathbf{r}) \in L^2(\Gamma_N)$. The functional
space for the Dirichlet  boundary term $\phi_0(\mathbf{r})$ can be
studied by the trace theorem \cite{Buffa2001}.
For the Lipschitz
boundary $\Gamma$, the function $\phi_0(\mathbf{r})$ can be extended
from $\phi(\mathbf{r})\in H^1(\Omega)$ to   $\phi_0(\mathbf{r})\in
H^{1/2}(\Gamma_D)$ by invoking the trace operator. In particular,
when $\Gamma_N=0$, the above equation shrinks to a Dirichlet
problem, which has the unique solution. However, when $\Gamma_D=0$,
it becomes a Neumann problem that is uniquely solvable
(up to a constant).

\subsection{Vector space decomposition}
Here, we state the well-known Helmholtz vector decomposition theorem~\cite{BladelEMbook} and introduce loop-tree decomposition which is commonly used in the computational electromagnetics (CEM) community~\cite{Wilton1981,Wu1994}.

\begin{theorem} (Helmholtz decomposition.) A vector field $\mathbf{f}(\mathbf{r})$ can be split into the form
\begin{equation}
\mathbf{f}(\mathbf{r})=\nabla\varphi+\nabla\times\mathbf{v}.
\end{equation}
The first term $\nabla\varphi$ is the irrotational (curl-free) part,
and the second term $\nabla\times\mathbf{v}$ is the solenoidal
(divergence-free) part.
\end{theorem}

Assume the vector field is living in a vector space $\mathcal V$. The Helmhotz theorem indicates
\begin{equation}\label{eq2.3}
\mathcal V= \mathcal V_{irr} \oplus \mathcal V_{sol}
\end{equation}
where $\mathcal V_{irr}$ is irrotational subspace and $\mathcal V_{sol}$ is solenoidal subspace. These two subspaces are orthogonal to each other.

\begin{figure}[h]
\centerline{\includegraphics[width=0.5\columnwidth,draft=false]{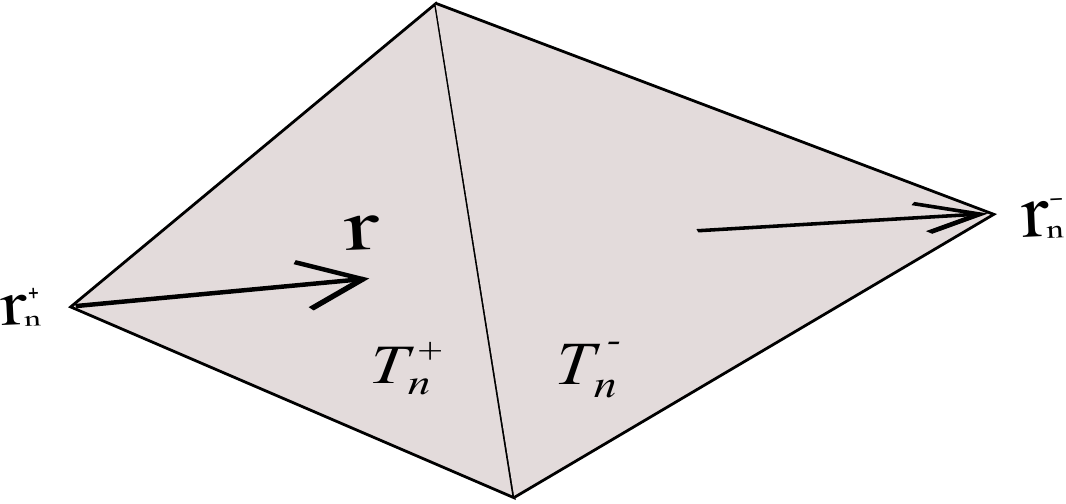}}
\caption{Rao-Wilton-Glisson functions.} \label{fig:rwg}
\end{figure}

\begin{definition} (Rao-Wilton-Glisson (RWG) functions~\cite{RWG}.) The
RWG function is defined on two adjacent triangles. The expression for
the expansion function is
\begin{equation}
\mathbf{\Lambda}_i(\mathbf{r})= \left\{
\renewcommand{\arraystretch}{1.5}
 \begin{array}{cl}
 \pm\frac{1}{2A_i^{\pm}}(\mathbf{r}-\mathbf{r}_i^{\pm}), &\mbox{ $\mathbf{r}\in T_i^{\pm}$}, \\
  0, &\mbox{ otherwise,}
 \end{array} \right.
\end{equation}
where $\pm$ denote the respective triangles, $\mathbf{r}_i^{\pm}$
and $A_i^{\pm}$ are the vertex points and areas of the respective
triangles,  and $T_i^{\pm}$ are the supports of the respective
triangles (refer to Figure \ref{fig:rwg}).
\end{definition}

Let $\mathcal T_h$ denotes a triangulation, on which we defined a set of RWG functions. Then these functions expand a space $\mathcal V(\mathcal T_h)$, namely
\begin{equation}
\mathcal V(\mathcal T_h)= span \{\mathbf{\Lambda}_1,\cdots,\mathbf{\Lambda}_{n_e}\},
\end{equation}
where $n_e$ is the number of RWG functions. In fact, $n_e$ amounts to the number of internal edges on triangulation $\mathcal T_h$.

\begin{definition} \label{def:lp}(Loop basis functions~\cite{VecchiLS}.)
A loop basis function is described by the surface curl of a vector function, namely,
\begin{equation}\label{eq2.8}
\mathbf{L}_i(\mathbf{r})=\nabla\times\hat{u}\,\sigma_i(\mathbf{r})
\end{equation}
where the scalar function $\sigma(\mathbf{r})$, also referred as
``solenoidal potential", is the linear Lagrange or
nodal interpolating basis. $\hat{u}$ stands for the unit normal vector of the simulation plane.
\end{definition}

\begin{figure}
  \centering
  \subfigure[A nodal interpolating basis function.]{
    \label{fig:loop:a} %% label for first subfigure
    \includegraphics[scale =0.4]{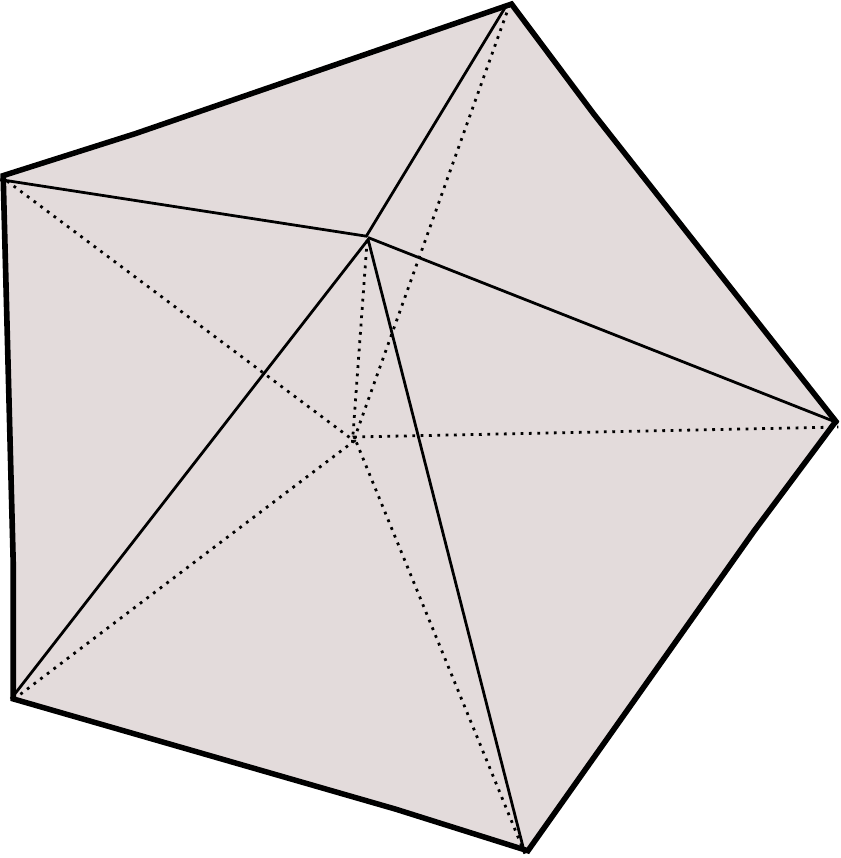}}
  \hspace{1in}
  \subfigure[A loop basis function.]{
    \label{fig:loop:b} %% label for second subfigure
    \includegraphics[scale =0.4]{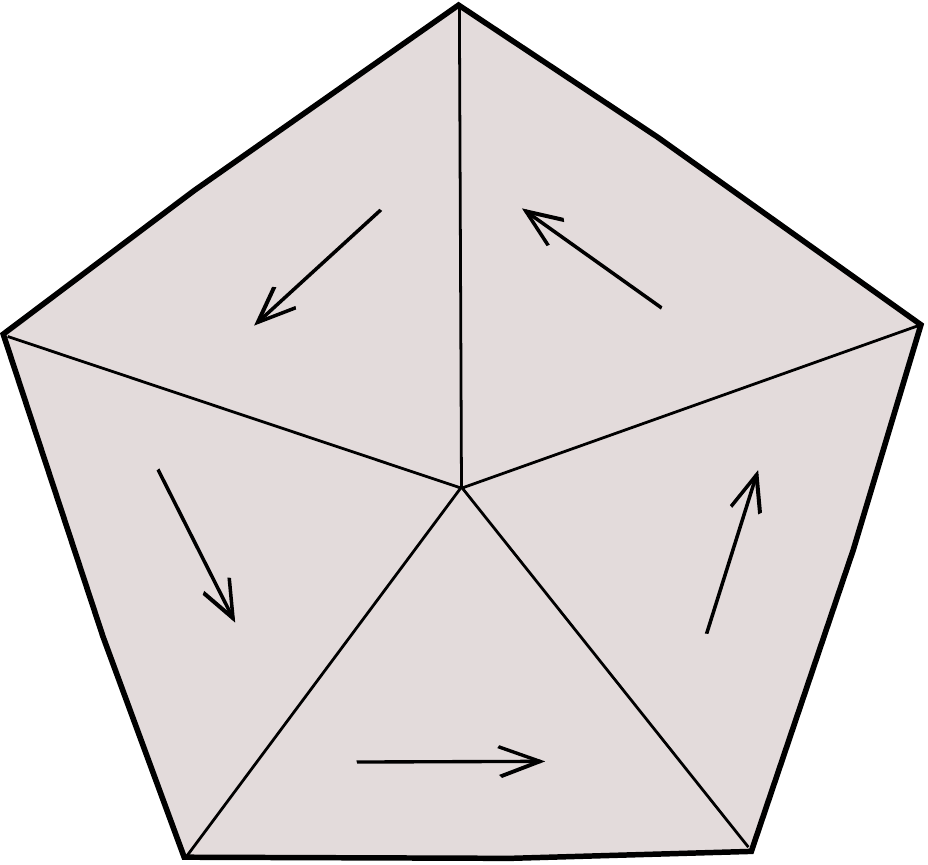}}
  \caption{Illustration of nodal basis and loop basis functions.}
  \label{fig:loop} %% label for entire figure
\end{figure}

Fig.~\ref{fig:loop:a} shows a typical nodal basis function, $\sigma_i(\mathbf{r})$,  which is a piecewise linear
function with support on the triangles that has a vertex at the $i$th node of the mesh, attaining a unit value at node $i$, and linearly approaching zero on all neighboring nodes. Moreover, Fig.~\ref{fig:loop:b} illustrates the loop basis function $\mathbf{L}_i$ associated with an interior node $i$. Within the triangles attached to node $i$, $\mathbf{L}_i$ has a vector direction parallel to the edge opposite to node $i$ and forms a loop around node $i$.

Since loop basis functions are divergence free, they expand the solenoidal space
\begin{equation}
 \mathcal V_{sol}(\mathcal T_h)= span \{\mathbf{L}_1,\cdots,\mathbf{L}_{n_l}\},
\end{equation}
where $n_l$ is the number of loop basis function that is the same as the number of inner nodes on $\mathcal T_h$.

\begin{definition}(Tree basis functions.) The tree basis consists of RWG functions that lie along a tree structure connecting the centroids of adjacent triangular patches.
\begin{equation}
\begin{array}{ll}
\mathbf{T}_i(\mathbf{r}) = \mathbf{\Lambda}_{k(i)}(\mathbf{r}),&\mbox{ if $\mathbf{\Lambda}_{k(i)}$ corresponds to tree edge},
\end{array}
\end{equation}
 where the number of tree basis functions $n_t$ equal the number of triangles minus one, $n_t= n_p-1$.
\end{definition}

\begin{figure}[h]
\centerline{\includegraphics[width=0.3\columnwidth,draft=false]{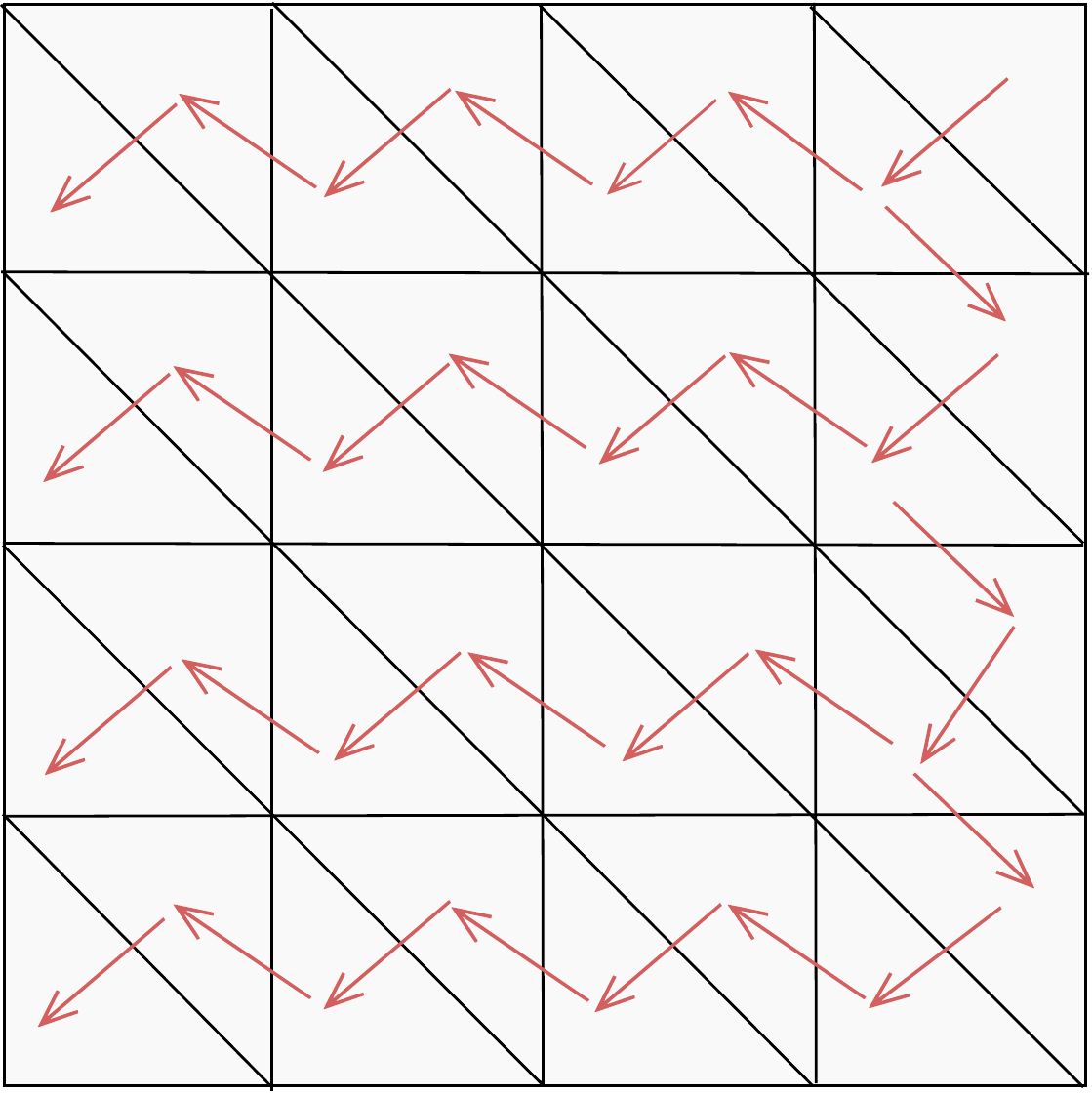}}
\caption{A possible tree basis illustration. Every arrow stands for
an RWG function that is the member of tree basis.} \label{fig:tree}
\end{figure}

For the structure shown in Fig.~\ref{fig:tree}, every edge corresponds to a RWG functions.  One possible choice of
the tree basis is illustrated, where those RWG functions corresponding to arrows consist of tree basis.
The tree basis functions have the property
$
\nabla\cdot\mathbf{T}_i(\mathbf{r})\neq 0
$. Hence, they expand the space $\mathcal V_t$.

It has been proven that loop basis functions and tree basis functions together span the same space as RWG functions, namely
\begin{equation}\label{eq2.9}
\mathcal V(\mathcal T_h)= \mathcal V_{sol}(\mathcal T_h)\cup\mathcal V_t(\mathcal T_h).
\end{equation}
It should be noted that the tree space $V_t(\mathcal T_h)$ is not identical to the irrotational space $V_{irr}(\mathcal T_h)$. Thus $V_t(\mathcal T_h)$ is not orthogonal to $\mathcal V_{sol}(\mathcal T_h)$. Eq.~(\ref{eq2.9}) is often called \emph{quasi-Helmholtz decomposition} in sense that it approximately accomplish Helmhotlz decomposition Eq.~(\ref{eq2.3}).

\begin{definition}(Pulse basis functions.) The pulse basis function is defined as
 \begin{equation}\label{eq2.10}
p_i(\v r)=\left\{
\begin{array}{ll}
1, & \v r \in \mbox{$i$-th patch},\\
0, & \mbox{otherwise}.
\end{array}\right.
\end{equation}
\end{definition}

We use pulse basis functions to expand the scalar quantity, such as charge $\rho$ and potential $\phi$. On a particular mesh $\mathcal T_h$, the pulse basis functions expand the space
\begin{equation}
\mathcal Q(\mathcal T_h)= span \{p_1,p_2,\cdots,p_{n_p}\},
\end{equation}
where $n_p$ is the number of patches. In some cases, due to charge neutrality condition, there are only $n_p-1$ basis functions are independent. Thus, we introduce the reduced space
\begin{equation}
\mathcal Q'(\mathcal T_h)= span \{p_1,p_2,\cdots,p_{n_p-1}\}.
\end{equation}
Then, we have
\begin{equation}
\begin{array}{ll}
\nabla\cdot \mathcal V(\mathcal T_h)= \mathcal Q'(\mathcal T_h),& \nabla\cdot \mathcal V_t(\mathcal T_h)= \mathcal Q'(\mathcal T_h).
\end{array}
\end{equation}

%-------------------------------------------------------------------
%----------------------------------------------------------------
%----------------------------------------------------------------
\section{Poisson Solver Through Loop-Tree Bases}

In this section, we give a self-contained description of loop-tree bases Poisson solver which has first proposed in~\cite{fpsnbc}.

As a model problem, we first consider the Neumann problem with homogeneous boundary condition, that is,
\begin{equation}\label{eq:npeq}
\begin{array}{ll}
\displaystyle \nabla\cdot\left(\epsilon_r(\mathbf{r})\nabla\phi(\mathbf{r})\right)=-\rho(\mathbf{r})/\epsilon_0 &\mbox{ for $\mathbf{r}\in\Omega$} \\
\frac{\partial}{\partial n}\phi(\mathbf{r})=0 &\mbox{ for $\mathbf{r}\in\Gamma$}
\end{array}
\end{equation}
where $\Gamma$ includes all the boundaries of $\Omega$.

Instead of solving Eq.~(\ref{eq:npeq}) directly, this novel method finds solutions by the following two sequential steps:

\begin{enumerate}

\item
    Find a $\v D(\v r)\in\mathcal V(\mathcal T_h)$ such that
    \begin{equation}\label{eq3.1}
    \nabla \cdot \v D(\v r)= \rho(\v r),
    \end{equation}
    and
    \begin{equation}\label{eq3.3}
    \nabla\times\frac{\v D(\v r)}{\epsilon(\v r)}= 0,
    \end{equation}
    with proper boundary conditions.

\item
    Find a $\phi(\v r)\in \mathcal Q(\mathcal T_h)$ such that
    \begin{equation}\label{eq3.4.0}
    -\nabla\phi(\v r)= \frac{\v D(\v r)}{\epsilon(\v r)}.
    \end{equation}
\end{enumerate}

The point of departure for this method is to expand the electric flux by
\begin{equation}\label{eq3.5}
\mathbf{D}(\mathbf{r})=\mathbf{D}_l(\mathbf{r})+\mathbf{D}_t(\mathbf{r})
\quad\in\mathcal V(\mathcal T_h)
\end{equation}
where
$$
\mathbf{D}_l(\mathbf{r})=\sum_{i=1}^{n_l} l_i\mathbf{L}_i(\mathbf{r})\quad\in \mathcal V_{sol}(\mathcal T_h)
$$
and
$$
\mathbf{D}_t(\mathbf{r})=\sum_{i=1}^{n_t}
t_i\mathbf{T}_i(\mathbf{r})\quad\in \mathcal V_t(\mathcal T_h).
$$

The way to obtain electric flux $\v D$ boils down to find these two subspace parts step by step.

\subsection{Tree space part}

Since the loop space part is divergence free, using Eq.~(\ref{eq3.4.1}) into (\ref{eq3.1}) leads to
\begin{equation}\label{eq3.5.0}
\begin{array}{ll}
    \nabla\cdot\mathbf{D}_t(\mathbf{r})=\rho(\mathbf{r}),&\mbox{ $\mathbf{r}\in\Omega$},\\
     D_n(\v r)= 0,&\mbox{ $\mathbf{r}\in\Gamma$},
\end{array}
\end{equation}
where $D_n(\v r)$ is the value of electric flux normal component on $\Gamma$. A Galerkin process solve the above problem. Testing the first equation of (\ref{eq3.5.0}) by a set of pulse basis function $\{p_1,p_2,\cdots,p_{n_p}\}$, we obtain the system
\begin{equation}\label{eq3.7.0}
K t= b
\end{equation}
where
\begin{equation}\label{eq3.8}
\begin{array}{ll}
K_{ij}= \left\langle p_i(\mathbf{r}),\nabla \cdot \textbf{T}_j(\mathbf{r}) \right\rangle,& b_i= \left\langle p_i(\mathbf{r}),\rho(\mathbf{r})\right\rangle.
\end{array}
\end{equation}

In the above, the reaction inner product between two functions is defined as
$$
\left\langle f_1, f_2 \right\rangle= \int f_1^\ast(\mathbf{r}) f_2(\mathbf{r}) d\mathbf{r}
$$
where the integral is assumed to converge. The matrix system (\ref{eq3.7.0}) can be solved with $\textit{O}(N_t)$
operations using the fast tree solver.

\subsection{Loop space part}

It is well known that the electric field, $\mathbf E$, is curl-free. This property is also implied in Eq.~(\ref{eq3.3}).
We can use a set of nodal basis function to test this equation:
\begin{equation}
\begin{array}{ll}
\langle \hat u\sigma_i(\v r), \nabla\times\frac{\v D(\v r)}{\epsilon(\v r)}\rangle=0,& i=1,2,\cdots,n_l.
\end{array}
\end{equation}
By using integrate by parts, the above becomes
\begin{equation}\label{eq3.10.0}
\begin{array}{ll}
\langle \nabla\times\hat u \sigma_i(\v r),\frac{\v D(\v r)}{\epsilon(\v r)}\rangle= 0,& i=1,2,\cdots,n_l.
\end{array}
\end{equation}
Hence, by using Eq.~(\ref{eq2.8}), we have
\begin{equation}\label{eq3.10}
\begin{array}{ll}
\langle \v L_i(\v r),\frac{\v D(\v r)}{\epsilon(\v r)}\rangle= 0,& i=1,2,\cdots,n_l.
\end{array}
\end{equation}
This equation indicates that the electric field is orthogonal to the loop space, namely,
$$
\mathbb{P}_L \mathbf E = 0,
$$
where $\mathbb{P}_L$ denotes the projection operator from the electric field $\v E$ space onto the loop space.

Furthermore, by using Eq.~(\ref{eq3.5}) and expanding $\mathbf{D}_l(\mathbf{r})=\sum_{i=1}^{n_l} l_i\mathbf{L}_i(\mathbf{r})$, Eq.~(\ref{eq3.10}) can be converted to the matrix form
\begin{equation}\label{eq3.11}
G l= c
\end{equation}
where
\begin{equation}\label{eq3.12}
\begin{array}{ll}
G_{ij}= \left\langle \v L_i(\mathbf{r}), \v L_j(\mathbf{r})/\epsilon(\v r) \right\rangle, & c_i= -\left\langle \v L_i(\mathbf{r}),\v D_t(\v r)/\epsilon(\v r)\right\rangle.
\end{array}
\end{equation}

Normally, commonly used iterative methods, such as BiCGSTB and GMRES~\cite{BiCGSTAB,gmres1986}, could be employed to solve Eq.~(\ref{eq3.11}). From our numerical experiments, this procedure dominates the whole computational cost.

\subsection{Obtain the potential}

Finding the potential amounts to solving Eq.~(\ref{eq3.4.0}). This can be done by expanding the potential a set of pulse basis function, $\phi(\mathbf{r})=\sum_{i=1}^{n_p}\nu_i p_i(\mathbf{r})$. Then a standard Garlerkin process can be used, with testing with tree basis functions, to achieve the following system
\begin{equation}\label{eq3.13}
K^t \nu= d
\end{equation}
where
\begin{equation}\label{eq3.14}
\begin{array}{ll}
K^t_{ij}= \left\langle\nabla \cdot \textbf{T}_i(\mathbf{r}), p_j(\mathbf{r}) \right\rangle, & d_i= \left\langle \v T_i(\mathbf{r}),\v D(\v r)/\epsilon(\v r)\right\rangle.
\end{array}
\end{equation}
The matrix $K^t$ is just the transpose matrix of $K$ in Eq.~(\ref{eq3.7.0}). One merit of this method is that the solution of Eq.~(\ref{eq3.13}) can be achieved by using the same fast tree solver because the del operator ($\nabla$) is the transpose of the divergence operator ($\nabla\cdot$).

\subsection{Other kinds of boundary conditions} There is a need to treat other kinds of boundary conditions. Here, we address the treatment for inhomogeneous Neumann boundary condition and Dirichlet boundary condition.

\subsubsection{Inhomogeneous Neumann boundary condition} Suppose the Neumann boundary condition in  Eq.~(\ref{eq:npeq}) is
\begin{equation}
\begin{array}{ll}
\frac{\partial}{\partial n}\phi(\mathbf{r})=g(\v r), & \mathbf{r}\in\Gamma.
\end{array}
\end{equation}
To guarantee the existence of solution, $g(\v r)$ should satisfy
\begin{equation}\label{eq3.16}
\int_{\Gamma} -\epsilon(\mathbf{r}) g(\mathbf{r}) d\textit{l}+ \int_{\Omega} \rho(\mathbf{r}) d\mathbf{r} =0.
\end{equation}

To impose the proper boundary condition, we need to modify the right hand side of Eq.~(\ref{eq3.8}) as
\begin{equation}\label{eq3.17}
b_i= \left\langle p_i(\mathbf{r}),\rho(\mathbf{r})\right\rangle-\int_{\Gamma_i} \epsilon(\mathbf{r}) g(\mathbf{r}) d\textit{l},
\end{equation}
if $p_i(\mathbf{r})$ is defined on a patch involving with boundary and $\Gamma_i$ stands for the boundary connecting this patch.

\subsubsection{Dirichlet boundary condition}

For the Dirichlet problems or mixed boundary problems, the specified Dirichlet boundary condition is posed by introducing a small region with high permittivity parameter.

As a model problem, we consider the problem showing in Fig.~\ref{fig:PE}. The governing equation is
\begin{equation}\label{s4eq0}
\renewcommand{\arraystretch}{1.5}
\begin{array}{ll}
\nabla\cdot\left(\epsilon_r(\mathbf{r})\nabla\phi(\mathbf{r})\right)=-\rho(\mathbf{r})/\epsilon_0, &\mbox{ for $\mathbf{r}\in\Omega$},   \\
\frac{\partial}{\partial n}\phi(\mathbf{r})=g(\mathbf{r}), &\mbox{ for $\mathbf{r}\in\Gamma_N$}, \\
\phi(\mathbf{r})=V_l, &\mbox{ for $\mathbf{r}\in\Gamma_{Dl}$}, \\
\phi(\mathbf{r})=V_r, &\mbox{ for $\mathbf{r}\in\Gamma_{Dr}$},
\end{array}
\end{equation}
where $V_l$ and $V_r$ are potential values imposed on left and right part of Dirichlet boundary, $\Gamma_{Dl}$ and $\Gamma_{Dr}$, respectively.

This problem can be broken into two parts. First, we need to solve the following equations
\begin{equation}\label{s4eq1}
\renewcommand{\arraystretch}{1.5}
\begin{array}{ll}
\nabla\cdot\left(\epsilon_r(\mathbf{r})\nabla\phi_1(\mathbf{r})\right)=-\rho(\mathbf{r})/\epsilon_0, &\mbox{ for $\mathbf{r}\in\Omega$}, \\
\frac{\partial}{\partial n}\phi_1(\mathbf{r})=g(\mathbf{r}), &\mbox{ for $\mathbf{r}\in\Gamma_N$}, \\
\phi_1(\mathbf{r})=V_l, &\mbox{  for $\mathbf{r}\in\Gamma_{Dl}$}, \\
\phi_1(\mathbf{r})=V'_r, &\mbox{  for $\mathbf{r}\in\Gamma_{Dr}$}, \\
\end{array}
\end{equation}
where $V'_r$ is a potential value arisen when we approximate this set equations.
Then, we can solve the second problems
\begin{equation}\label{s4eq6}
\renewcommand{\arraystretch}{1.5}
\begin{array}{ll}
\nabla\cdot\left(\epsilon_r(\mathbf{r})\nabla\phi_2(\mathbf{r})\right)=0, &\mbox{ for $\mathbf{r}\in\Omega$}, \\
\frac{\partial}{\partial n}\phi_2(\mathbf{r})=0, &\mbox{  for $\mathbf{r}\in\Gamma_N$},   \\
\phi_2(\mathbf{r})=0, &\mbox{  for $\mathbf{r}\in\Gamma_{Dl}$}, \\
\phi_2(\mathbf{r})=V_r-V'_r, &\mbox{ for $\mathbf{r}\in\Gamma_{Dr}$}.
\end{array}
\end{equation}
Obviously, the solution of original problem is just
\begin{equation}\label{s4eq7.3}
\phi(\mathbf{r})=\phi_1(\mathbf{r})+\phi_2(\mathbf{r}).
\end{equation}

A thorough description of handling these two problems can be found in~\cite{myJCP}. The repetitious details need not be given here.

\subsection{Comparison with nodal basis FEM}
The Poisson solver through loop-tree basis has demonstrated its efficiency in~\cite{myJCP}. Here, we further compare the difference between the proposed method and traditional nodal basis FEM in terms of basis function space.

Let $\mathcal U(\mathcal T_h)$ be the space spanned by nodal basis function defined on mesh $\mathcal T_h$, which is also called \emph{node-type subspace}. The nodal basis FEM seeks a solution $\phi\in \mathcal U(\mathcal T_h)$.
Moreover, there is a relation
\begin{equation}
\begin{array}{ll}
\nabla\mathcal U(\mathcal T_h)\subset \mathcal Y(\mathcal T_h),& \nabla\times\mathcal Y(\mathcal T_h)= \mathcal V(\mathcal T_h),
\end{array}
\end{equation}
where $\mathcal Y(\mathcal T_h)$ refers to the space spanned by edge-element basis functions. This is also the relation of Whitney $0, 1 ,2$ form~\cite{Whitney1988}.

From the above, it is clear that the electric field space obtained from traditional nodal basis FEM, $\nabla\mathcal U(\mathcal T_h)$, is a subset of $\mathcal Y(\mathcal T_h)$. A function of $\nabla\mathcal U(\mathcal T_h)$ defined on a particular triangular patch is just a constant vector. On the other hand, the electric flux of the proposed method is expressed by functions in  $\mathcal V(\mathcal T_h)$, which are 2-form basis functions. These functions have normal continuity property and are natural to represent the electric flux field in electromagnetics. In other words, given the same mesh $\mathcal T_h$, the obtained electric field is more accurate since it use functions in $\mathcal V(\mathcal T_h)$. This is beneficial for some applications.

%-----------------------------------------------------------------
%-----------------------------------------------------------------
%-----------------------------------------------------------------

\section{Multilevel Method}

As mentioned in the above section, finding the loop space part plays a critical role within solution procedures. This iterative process dominates the computational time of the whole solution. On the other hand, it is observed that the convergent speed becomes slow when higher accuracy is required. Therefore, it is imperative to find an approach to accelerate convergent procedure. In this section, a multilevel method (also known as multiresolution method) though hierarchical loop basis will be presented.

\subsection{Nested Mesh and Hierarchical Loop Basis}

\begin{figure}[h]
\centerline{\includegraphics[scale=0.35]{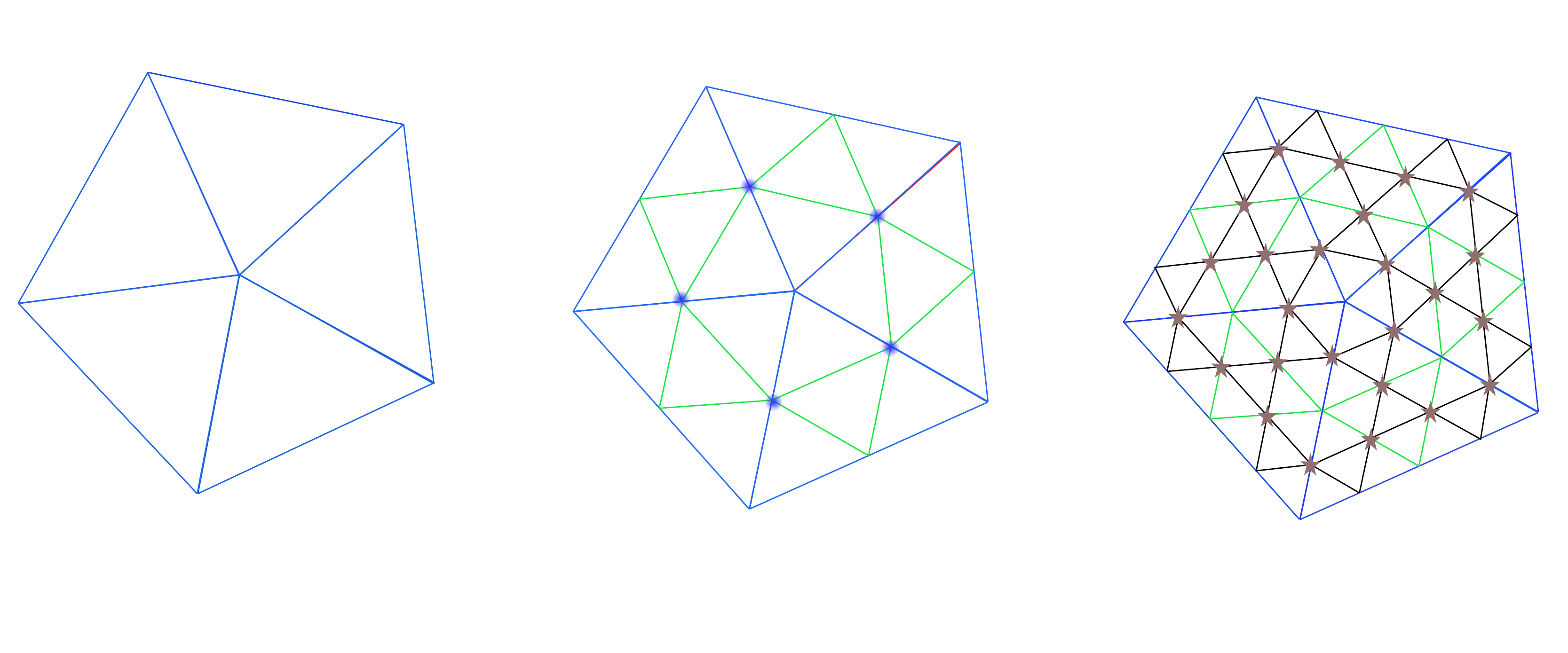}}
\caption{The illustration of nested mesh scheme. Left: level-$0$; Middle:level-$1$; Right: level-$2$. } \label{fig:nestM}
\end{figure}

First, we introduce a nested mesh scheme, in which several meshes of different granularity are utilized.  The coarsest mesh is generated first. We label this coarsest mesh with level-$0$ mesh associated with triangulation $\mathcal T_0$ and use $\mathcal U_0$ to denote the space spanned by nodal basis functions defined on it. Then higher level mesh are subsequently built by subdividing each triangular element of the mesh used in the previous level into four equal-area sub-triangles, as shown in Fig.~\ref{fig:nestM}. Let $\mathcal T_0$,$\mathcal T_1$,$\mathcal T_2$,... be the nested triangulations. Thus, on each mesh level we can define a nodal basis space, obtaining spaces $\mathcal U_0$, $\mathcal U_1$, $\mathcal U_2$,...

Suppose the finest level is level-$K$ mesh, it is evident that these spaces satisfy
\begin{equation}\label{eq4.1}
\mathcal U_0\subset \mathcal U_1\subset \mathcal U_2 \cdots \subset \mathcal U_K.
\end{equation}
Following the convention of \cite{Deuflhard1989}, the space $\mathcal V_{i+1}$ can be split into two parts
\begin{equation}\label{eq4.2}
\mathcal U_{i+1}= \mathcal U_i \oplus \mathcal W_{i+1}.
\end{equation}
In the above, functions in $\mathcal W_{i+1}$ are those nodal functions corresponds to new nodes added in the level-($i+1$) mesh after doing bisection. For example, in Fig.~\ref{fig:nestM}, those nodal basis functions corresponding to points marked by dots, which are defined on triangulations of level-$1$ mesh, are elements of $\mathcal W_1$. Similarly, those nodal basis functions of level-$2$ mesh corresponding to points marked by stars belong to the space $\mathcal W_2$.
Consequently, the space $\mathcal U_K$ satisfies this decomposition relation
\begin{equation}\label{eq4.3}
\mathcal U_K= \mathcal U_0 \oplus \mathcal W_1 \oplus\mathcal W_2\cdots\oplus\mathcal W_K.
\end{equation}
This implies a hierarchical nodal basis functions can be constructed in the following scheme. Assume level-0 triangulation $\mathcal T_0$ has $n_0$ nodes, we define nodal basis functions $\{\sigma_1,\sigma_2,\cdots,\sigma_{n_0}\}$ at these nodes, which expand the space $\mathcal U_0$. At the next level, it creates a triangulation $\mathcal T_1$ have $n_1$ nodes by refinement process, thereby defining nodal basis functions $\{\sigma_{n_0+1},\sigma_{n_0+2},\cdots,\sigma_{n_1}\}$ only at the $n_1-n_0$ nodes added at this level, which expand the space $\mathcal W_1$ . Processing this refinement until there are $K$ levels, we obtain the hierarchical nodal basis functions $\{\sigma_1,\sigma_2,\cdots,\sigma_{n_K}\}$.
%This implies that hierarchical multilevel basis functions in $\mathcal W_i$, $i=1,\cdots,K$ could be constructed so that they span the same space with the original nodal basis function space $\mathcal V_K$.

Once the hierarchical nodal basis functions are found, the corresponding hierarchical loop basis functions are obtained from definition~\ref{def:lp}:
$$
\mathbf L(\mathbf r)= \nabla\times \hat{u}\sigma(\mathbf r).
$$
The above is applicable for any mesh level.

\subsection{Multilevel System}

Considering the finest level-$K$ mesh, in normal loop basis representation, the equivalent linear system of  (\ref{eq3.11}) is
\begin{equation}\label{{eq4.5}}
G_K l_K= c_K.
\end{equation}

The use of hierarchical loop basis yields
\begin{equation}\label{eq4.6}
\tilde{G}_K \tilde{l}_K= \tilde{c}_K.
\end{equation}
where
\begin{equation}\label{eq4.7}
\begin{array}{l}
\tilde{G}_K = S_K^T G_K S_K,\\
\tilde{c}_K= S_K^T c_K,\\
l_K= S_K \tilde{l}_K.
\end{array}
\end{equation}
In the above, $S_K$ is the transformation matrix from the original nodal basis to hierarchical basis. By doing so, the resulting linear system involves information on all $K$ levels meshes, which renders iterative solvers more efficient. In addition, it is worth to note that we also call the multilevel method multilevel preconditioning since it actually perform as a preconditioner as can be seen from Eqs.~(\ref{eq4.7}).

The application of hierarchical nodal basis in FEM showed that the condition numbers of preconditioned stiffness matrix has a condition number behaving like $O\left(\left(\log \frac{1}{h}\right)^2\right)$ ( $h$ is mesh size), which is contrary to $O\left(\left(\frac{1}{h}\right)^2 \right)$ for normal nodal basis. Furthermore, the method of conjugate gradients needs only $O(N\log N)$ computational operations to reduce the energy of the error by a given factor~\cite{Yserentant1986347}. A simple interpretation can account for this improvement: Since the differential equation operator gives rise to a sparse matrix system which corresponds to near-neighbor interactions, each matrix-vector product will send the information $O(1)$ grid points away; Therefore, it takes $O(N^{0.5})$ steps to send the information completely through the simulation region in 2D problems \cite{ChewBook2001}; This situation is changed by introducing the hierarchical loop basis, where large loop functions as well as small ones co-exist; As a result, information can traverse the simulation domain quickly. This acceleration of convergence is also achieved if one use hierarchical loop basis or related preconditioning procedures.

%-----------------------------------------------------------------
%-----------------------------------------------------------------
%-----------------------------------------------------------------

\section{Numerical Results}

In this section, we will use the proposed multilevel Poisson solver to find solutions of several 2-D Poisson problems. The new algorithm described above has been implemented in C++ platform with Intel compiler. Moreover, all simulations listed below are performed on an ordinary laptop with the $2.66$ GHz CPU, $4$ GB memory, and Windows
operating system.

%\subsection{Simple heterogeneous Poisson problem}
\emph{Example 1 (Simple heterogeneous Poisson problem)}. In our first example, we consider a 2-D Poisson problem in a heterogeneous medium:
$$
\begin{array}{cl}
\nabla\cdot\left(\epsilon_r(x,y)\nabla\phi(x,y)\right)=-\pi \cos(\pi x)-\pi \cos(\pi y), & \mbox{ $\qquad$ $(x,y)\in\Omega$}
\end{array}
$$
with homogeneous Neumann boundary condition. The computational domain $\Omega$ is given by
$$
\Omega = \left[ 0,1 \right] \times \left[ 0,1 \right]
$$
and the relative permittivity is
$$
\epsilon_r=\left\{
\begin{array}{cl}
1, & x<0.5, \\
2, & x\geq0.5.
\end{array}\right.
$$

We further assume that a reference potential $2/\pi$ imposed at
the origin, then the problem has the close form solution
$$
\phi(x,y)= \frac{\cos(\pi x)+\cos(\pi y)}{\pi \epsilon_r(x,y)}.
$$

\begin{figure}
  \centering
  \subfigure[$x$ component of the electric flux density $\mathbf{D}$.]{
    \label{fig:subfig:a} %% label for first subfigure
    \includegraphics[scale =0.6]{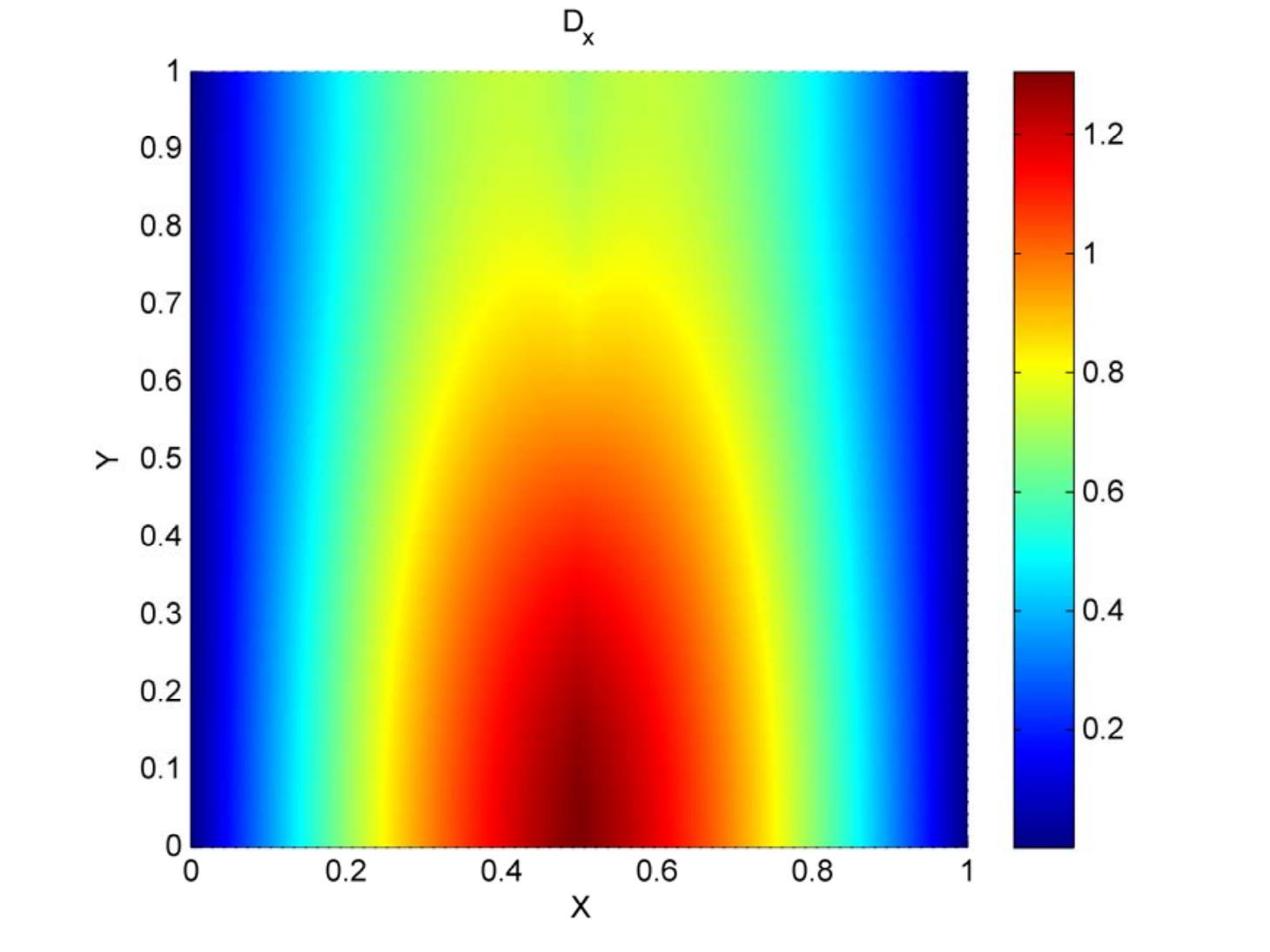}}
  \hspace{1in}
  \subfigure[$y$ component of the electric flux density $\mathbf{D}$.]{
    \label{fig:subfig:b} %% label for second subfigure
    \includegraphics[scale =0.6]{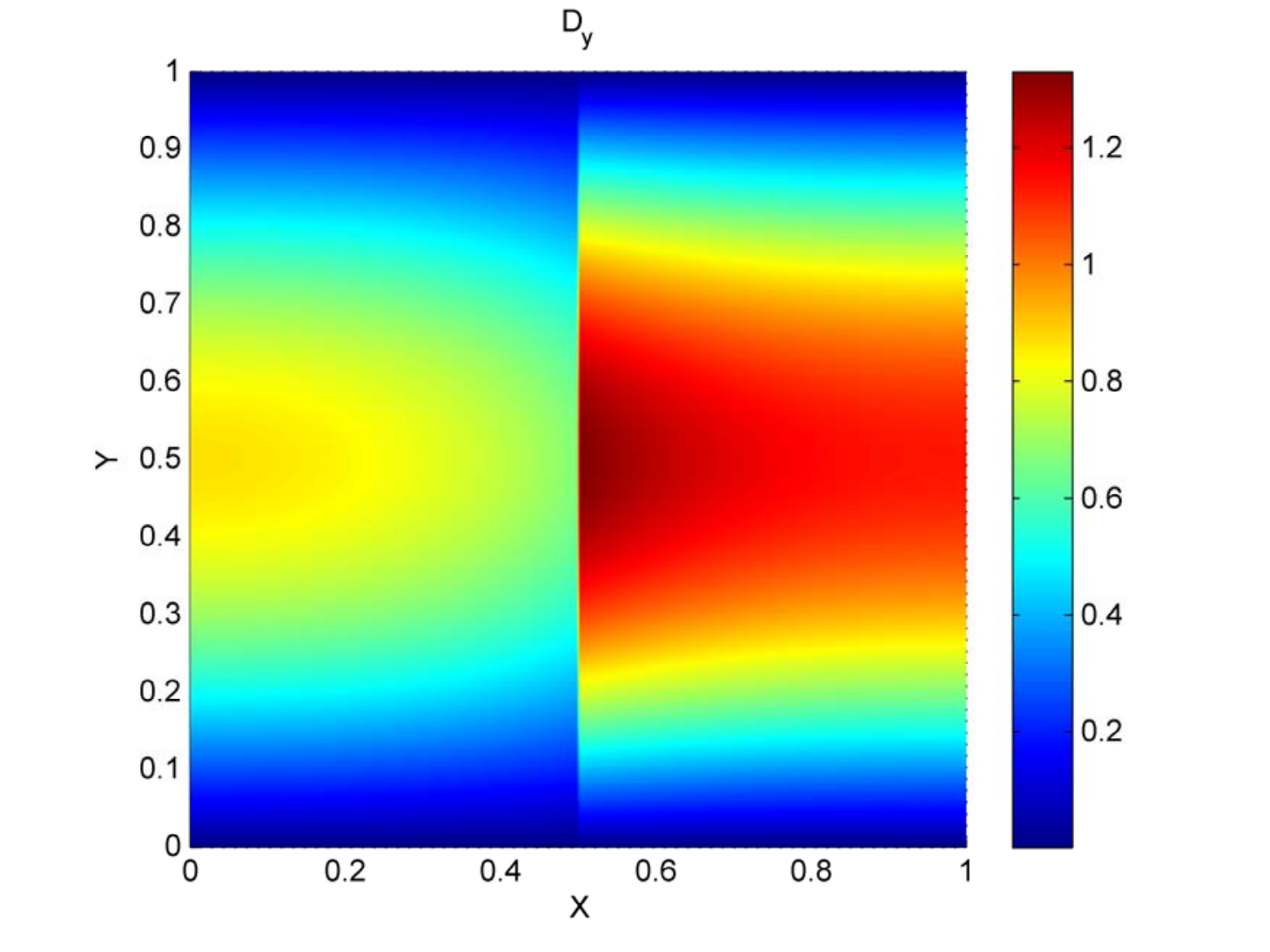}}
  \caption{The electric electric flux density calculated by the proposed method.}
  \label{fig:ex1} %% label for entire figure
\end{figure}

To validate the correctness of our code, we perform the bisection refinement procedure to obtain  $367$,$552$ triangular patches and solve the resulting discrete problem by GMRES method using the zero initial and stopping criterion $\delta<1\times10^{-5}$. Fig.~\ref{fig:ex1} shows the calculated electric flux. Because there is discontinuity
for $\epsilon_r$ at $x=0.5$, $y$-components, $D_y$, appears as an
abrupt change in the middle correspondingly, which is in complete
agreement with the fundamental boundary conditions of electromagnetism.

\begin{figure}[h]
\centerline{\includegraphics[scale=0.6]{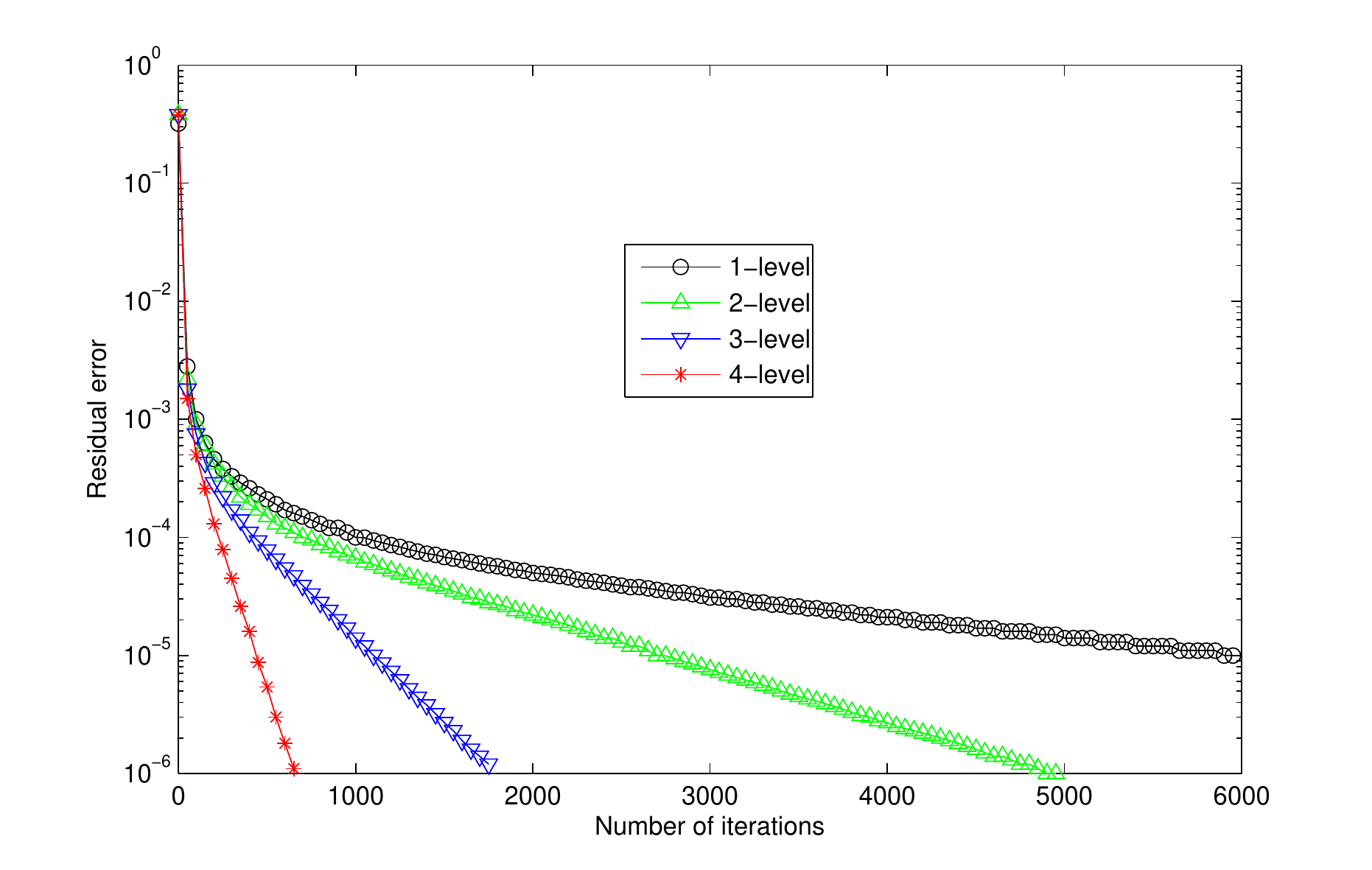}}
\caption{History of convergence.} \label{fig:conv}
\end{figure}

To study the effect of multilevel preconditioning, we further use the proposed method to solve the above problem with different levels of hierarchical loop basis functions. In this case, we use the same discretized mesh (with $1$,$470$,$208$ triangle patches). Fig.~\ref{fig:conv} shows the history of convergence. From this figure, we can see that the multilevel method through hierarchical loop basis functions improve convergent behavior dramatically. Moreover, the multilevel method converges much faster as more levels of hierarchical loop basis functions are used. On the contrary, previous Poisson solver through normal loop basis functions (corresponding to level-$1$) fails to convergent to $1\times10^{-5}$ even after $6000$ steps of iterations.

%\subsection{Complex Poisson problem with Dirichlet boundary condition}

\emph{Example 2 (A Poisson problem in irregular region.)} Next, we simulate the following two-dimensional Poisson equation
\begin{equation}\label{s5eq2}
\begin{array}{cl}
\nabla\cdot\left(\epsilon_r(\mathbf{r})\nabla\phi(\mathbf{r})\right)=-\delta(\mathbf{r}-\mathbf{r}'), &\mbox{ $\qquad$ for $\mathbf{r}\in\Omega$,}\\
\end{array}
\end{equation}
with boundary conditions
$$
\renewcommand{\arraystretch}{1.5}
\begin{array}{ll}
\phi(\mathbf{r})=1.0, &\mbox{ $\mathbf{r}\in\Gamma_1$,} \\
\phi(\mathbf{r})=0.8, &\mbox{ $\mathbf{r}\in\Gamma_2$,} \\
\frac{\partial}{\partial n}\phi(\mathbf{r})=0, &\mbox{ $\mathbf{r}\in$ other boundaries,}
\end{array}
$$
where $\mathbf{r}'$ is the point $(-0.2,0.6)$.

\begin{figure}[h]
\centerline{\includegraphics[scale=0.3,draft=false]{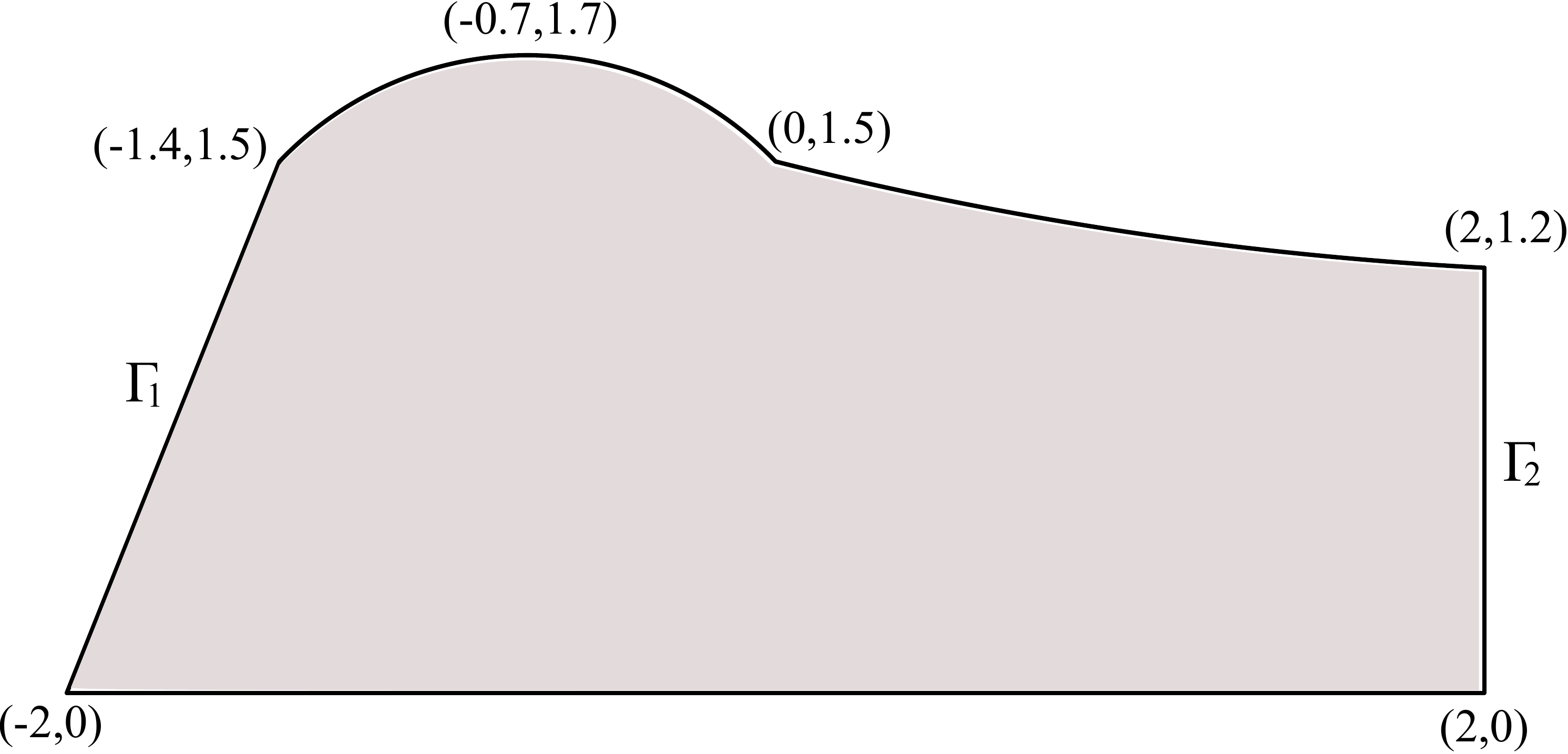}}
\caption{A two dimensional region where the Poisson problem is
defined.} \label{fig:flagshape}
\end{figure}

Fig.~\ref{fig:flagshape} shows the specifications of solution region
$\Omega$. This problem is excited by a line source that is located
at point $\mathbf{r}'$ with unit charge and imposed Dirichlet
boundary conditions on the left and right edges. In our simulation, we use the
following function to approximate the line source
$$
\delta(x,y)=\left\{
\renewcommand{\arraystretch}{1.5}
\begin{array}{cl}
625, & |x|<0.02, |y|<0.02, \\
0,   & \mbox{otherwise}.
\end{array}\right.
$$

\begin{figure}[h]
\centerline{\includegraphics[scale=0.6,draft=false]{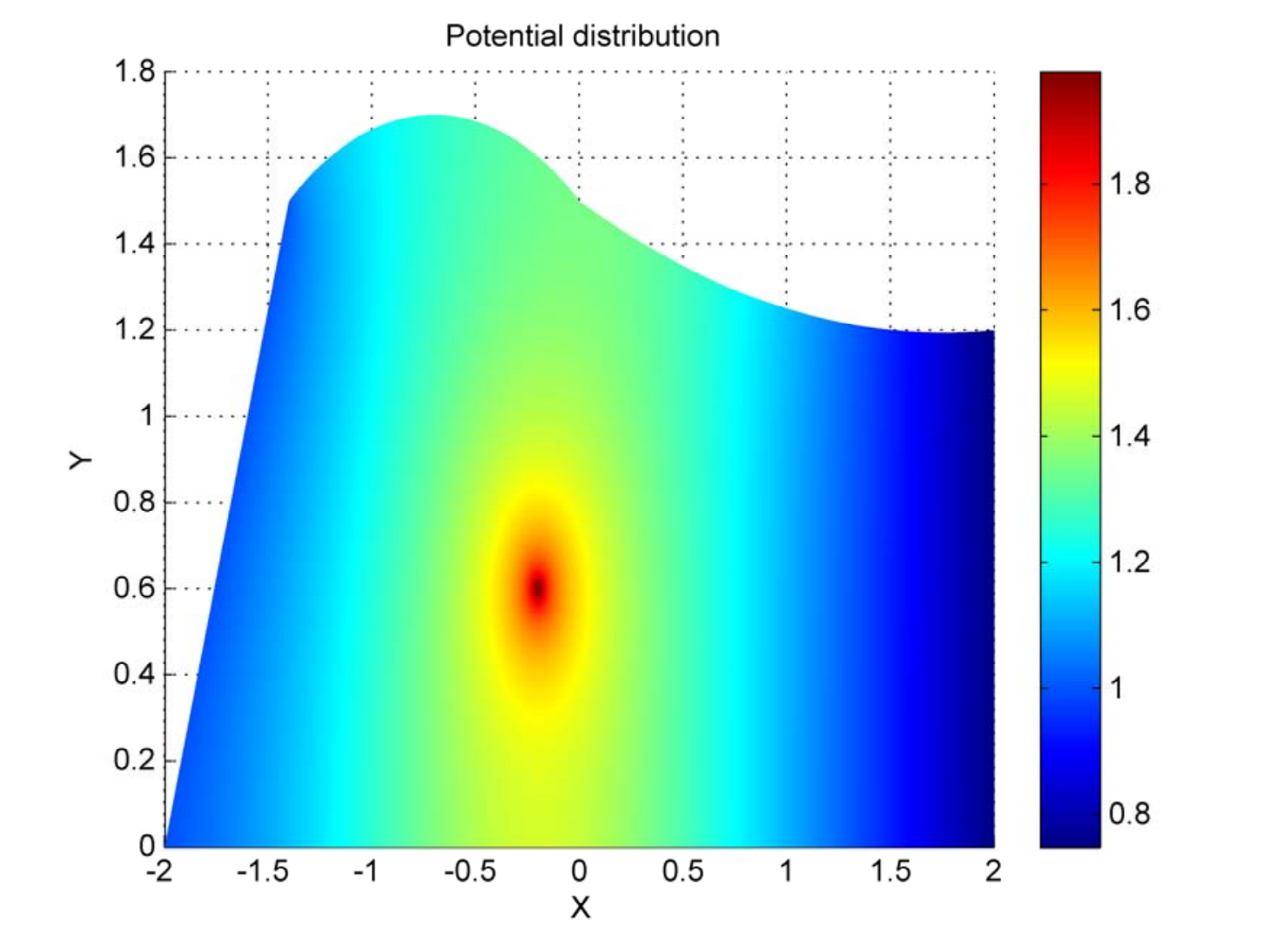}}
\caption{Calculated potential distribution.} \label{fig:ex2potl}
\end{figure}

Fig.~\ref{fig:ex2potl} illustrates the potential distribution calculated by the proposed method. This result agrees well with conventional FEM. Next, we examine the efficiency of proposed method compared with the Poisson solver through plain loop-tree basis as we increase the levels of hierarchical loop basis. At the start, the simulation domain is discretized by $33$,$083$ triangular patches, which corresponds to $16$,$264$ loop basis functions in 1-level system.  Table~\ref{tab:tab1} lists the iteration steps and the whole solution time with two different stopping criterions when hierarchical loop bases of different level are used. As can be seen from this table, the proposed method remarkably improve the efficiency compared with the previous method that use normal loop-tree basis. Because the transformation matrix $S$ is rather simple and sparse, which only depends on the triangulation. The workload of computing the product of the hierarchical basis matrix with a vector is nearly the same as that of computing the product of original matrix with a vector. Furthermore, since the iteration number does not depend on unknowns number $N$, this new method provides a multigrid speed of convergence. Thus, in our implementations, the required computational operations is of $O(N\log N)$ at most.

\begin{table}
\begin{center}
    \begin{tabular}{crclrclr}
    \hline
    &&&\multicolumn{2}{c}{proposed method}&&\multicolumn{2}{c}{previous method}\\
    \cline{4-5} \cline{7-8}
    & $N_l$ & &  iterations  & time(s) & &  iterations  & time(s)\\
    \hline
    \multicolumn{2}{c}{$1\times10^{-3}$}\\
    \cline{1-2}
    1-level & 16,264   & & 68  & 0.797 & &  &    \\

    2-level & 65,610 & & 80  & 3.500 & & 107  & 4.423   \\

    3-level & 263,551 & & 66  & 14.06 & & 107  & 18.835   \\

    4-level & 1,056,429 & & 70  & 63.063 & & 145  & 104.059   \\
    \hline
    \hline
    \multicolumn{2}{c}{$1\times10^{-4}$}\\
    \cline{1-2}
    1-level & 16,264 & & 162  & 1.703 & &  &    \\

    2-level & 65,610 & & 167  & 7.563 & & 267  & 11.012   \\

    3-level & 263,551 & & 161  & 32.095 & & 434  & 76.11   \\

    4-level & 1,056,429 & & 149  & 134.71 & & 948  & 725.5   \\
    \hline
    \end{tabular}
    \caption{A comparison between proposed method and previous method through loop-tree basis at the stopping criterions $1\times10^{-3}$ and $1\times10^{-4}$.}
    \label{tab:tab1}
\end{center}
\end{table}

%\begin{figure}[h]
%\centerline{\includegraphics[scale=0.6,draft=false]{memory.pdf}}
%\caption{Memory consumption versus the number of unknowns.} \label{fig:mem}
%\end{figure}

The memory consumption relies on the storage cost of two matrices: $S_K$ and $G_{l,K}$ in Eq.~(\ref{eq4.7}). As mentioned above, the transformation matrix $S_K$ only depends on the triangulation. Thus, it is possible to find its value from the triangulation data or store it in a highly sparse form. Then, the memory consumption is dominated by storage of $G_{l,K}$, which is of $O(N)$ complexity.

%------------------------------------------------------------
%------------------------------------------------------------
%------------------------------------------------------------

\section{Conclusion}

A new multilevel method through hierarchical loop basis functions for solving Poisson's equation resulting from electrostatic analysis is developed in this paper. This method is based on quasi-Helmholtz decomposition and termed hierarchical loop basis based Poisson Solver (hieLPS). The hierarchical loop basis is defined on a nested hierarchical mesh with corresponding bisection refinement scheme on  a triangular mesh. By first constructing hierarchical nodal basis function, the hierarchical loop basis functions are obtained by taking the surface curl of corresponding loop basis functions. The results show that this multilevel method serves as a good preconditioning for previous Poisson solver through normal loop-tree bases. If high level of hierarchical loop basis functions are used, iteration number could be reduced noticeably. The required computational cost of the new method is of $O(N\log N)$ at most, while the memory consumption is close to $O(N)$. This new method can be an alternative to the multilevel multigrid method.

\section*{Acknowledgments}
This work was supported in part by the Research Grants Council of
Hong Kong (GRF 711609, 711508, 711511 and 713011), in part by the
University Grants Council of Hong Kong (Contract No. AoE/P-04/08)
and HKU small project funding (201007176196).

%\bibliographystyle{siam}
%\bibliography{MR_fps}

\end{document}